\documentclass[aps,pra,twocolumn,superscriptaddress,longbibliography]{revtex4-1}

\usepackage[utf8]{inputenc}
\usepackage[english]{babel}

\usepackage{mathtools}
\usepackage{physics}
\usepackage{braket}
\usepackage{bbm}
\usepackage{bm}
\usepackage{amsbsy}
\usepackage{amsthm}
\usepackage{amssymb}
\usepackage{amsfonts}
\usepackage{amsmath}
\usepackage{dsfont} 
\usepackage{graphicx} 
\usepackage{epsfig}
\usepackage{epstopdf}
\usepackage{dsfont}
\usepackage{multibib}
\usepackage{color}
\usepackage{verbatim}
\usepackage{nomencl}
\usepackage[colorlinks]{hyperref}
\usepackage{float}

\newtheorem{defi}{Definition}

\newcommand{\M}[1]{\mathcal{#1}}

\newcommand{\ch}[1]{\left\{ { #1} \right\}}

\begin{document}
	
	\title{Genuine Multipartite Correlations Distribution in the Criticality of Lipkin-Meshkov-Glick Model}

	\author{Ant\^onio C. Louren\c{c}o}
	\email[]{lourenco.antonio.c@gmail.com}
	
	\affiliation{Departamento de F\'isica, Universidade Federal de Santa Catarina, CEP 88040-900, Florian\'opolis, SC, Brazil}
	
	\author{Susane Calegari}
	\affiliation{Departamento de F\'isica, Universidade Federal de Santa Catarina, CEP 88040-900, Florian\'opolis, SC, Brazil}
	
	\author{Thiago O. Maciel}
	\affiliation{Departamento de F\'isica, Universidade Federal de Santa Catarina, CEP 88040-900, Florian\'opolis, SC, Brazil}
	
	\author{Tiago Debarba}
	\affiliation{Departamento Acad{\^ e}mico de Ci{\^ e}ncias da Natureza, Universidade Tecnol{\'o}gica Federal do Paran{\'a} (UTFPR), Campus Corn{\'e}lio Proc{\'o}pio, Avenida Alberto Carazzai 1640, Corn{\'e}lio Proc{\'o}pio, Paran{\'a} 86300-000, Brazil.}
	
	\author{Gabriel T. Landi}
	\affiliation{Instituto de F\'isica, Universidade de S\~ao Paulo, CEP 05314-970, S\~ao Paulo, S\~ao Paulo, Brazil}
	
	\author{Eduardo I. Duzzioni}
	\affiliation{Departamento de F\'isica, Universidade Federal de Santa Catarina, CEP 88040-900, Florian\'opolis, SC, Brazil}

	
	
	\begin{abstract}
		Correlations play a key role in critical phenomena. 
		Quantities such as the entanglement entropy, for instance, provide key insights into the mechanisms underlying quantum criticality. 
		Almost all of our present knowledge, however, is restricted to  bipartite correlations. 
		Some questions still remain unanswered, such as: What parcel of the total correlations are genuinely $k$-partite? With the goal of shedding light on this difficult question, in this paper we put forth a detailed study of the behavior of genuine multipartite correlations (GMC) of arbitrary orders in the Lipkin-Meshkov-Glick model.
		We find that  GMC of all orders serve to signal the second order quantum phase transition presented in the  model. Applying finite-size scaling methods, we also find the critical exponents for some orders of correlations.
	\end{abstract}
	
	
	\maketitle
	
	\section{Introduction}
	
	Phase transitions emerge from the complex correlations developed between the microscopic constituents. 
	Understanding and characterizing these correlations has, therefore, always been a central problem in statistical physics.
	This is particularly more so for quantum phase transitions, for which one may employ  concepts from quantum information theory. 
	For example, the divergence of the entanglement entropy as one crosses the critical point is related to the underlying conformal theory that dictates the universal properties of a quantum phase transition \cite{Vidal2003}. 
	This has led to several studies aimed at characterizing entanglement in a variety of different critical systems  \cite{iemini20152,vidal2006,dusuel2004,dusuel2005,latorre2005}, including the first direct experimental measurement of the entanglement entropy in a superfluid \cite{Islam2015}. 
	
	Most of our present knowledge on this subject, however, is restricted to bipartite correlations. 
	The extension to a multipartite scenario is highly non-trivial, for two main reasons. 
	The first is related to the factorial large number of partitions that one can divide a system comprised of $N$ parties, making the problem difficult to analyze.
	The second is related to the difficulties in constructing measures of
	\emph{genuine} multipartite correlations (GMC) \cite{bennett2011}. 
	In a system with $N$ parts, a genuine correlation of order $k\leq N$ represents the total amount of correlations that cannot be obtained from  clusters of size smaller than $k$. 
	Hence, GMC should be able to quantify what part of the total correlations is distributed between clusters of different sizes. 
	This has applications in e.g., dimmerization in aperiodic spin chains \cite{Vieira2005} or the formation of strings in the Fermi-Hubbard model \cite{Chiu2018}. 
	GMC can therefore be a valuable tool in our understanding of quantum criticality. 
	
	The characterization of multipartite correlations in quantum critical systems has thus far focused almost exclusively in multipartite entanglement, which has been explored in a variety of models \cite{DeOliveira2006,DeOliveira2006a,Anfossi2005,Sun2014,Vidal2003}. 
	The current available measures of multipartite entanglement, however, are either ill defined or too complex to be computed  \cite{Horodecki2009} (for a review on approaches to characterize multipartite entanglement in many-body systems, see \cite{Amico2008}). For this reason, such studies still remain scarce.  
	
	More recently, Girolami \emph{et al.} have put forth a formalism for computing GMC which relies only on knowledge of the quantum relative entropy (Kullback-Leibler divergence) \cite{girolami2017}. 
	The formalism accounts for both quantum and classical correlations and is based on general distance-based concepts, formalized in Ref.~\cite{modi2010unified}, thus making it much more tractable.
	This framework has since been applied to GHZ \cite{susa2018} as well as Dicke states \cite{calegari2019}.
	
	In this work, we calculate genuine $k$-partite correlations in the ground state of the Lipkin-Meshkov-Glick (LMG) model by the framework presented in Ref. \cite{girolami2017}. Also, we show how is the distribution of correlations for a system of many particles and that these correlations signal the already known second order quantum phase transition (QPT). 
	In addition, we use a method of finite-size scaling (FSS) to find the critical exponent of some orders $k$ of correlations with emphasis on the total correlation, the bipartite correlation, and the tripartite correlation.     
	
	This work is organized as follows: In Sec. \ref{sec:mat} is presented the materials and methods, where we introduce the measures used to calculate the genuine $k$-partite correlations, the LMG model, its QPT, and the FSS theory. The analysis of the genuine $k$-partite correlations, the verification that all orders of $k$ signal the second order QPT, and the achievement of the critical exponents via FSS for total correlation, bipartite correlation and tripartite correlation, are presented in Sec. \ref{sec:res}. Finally, the conclusions and perspectives are left for Sec. \ref{sec:conclusion}.
	
	\section{Materials and Methods}
	\label{sec:mat}
	\subsection{Measures of genuine $k$-partite correlations}
	
	Consider an $N$-partite system described by the density matrix $\rho_N\in \M{D}(\M{H}_N = \M{H}_{[1]}\otimes \ldots \otimes \M{H}_{[N]})$, where each partition 
	$\rho_{j} = \tr_{N/j}{( \rho_N)  } $ is the state of the subsystem $j$, where $\tr_{N/j}$ indicates the trace over all partitions except $j$, such that $\rho_{j} \in \M{D}(\M{H}_{[j]})$. It is important to emphasize that each partitioning $\M{H}_{[j]}$ can also be a multipartite system, indeed the 
	number of sub partitions in each subsystem will be useful to define the \emph{genuine} correlations. Now, let us consider that the system has $m$-partitions $\ch{\M{H}_{k_j}}_{j = 1}^{m}$ and $k_j$ denotes the number of partitions in each subsystem 
	\begin{equation}
	S_{k_j} = \ch{ S_{[1]},\ldots, S_{[k_j]} }, \quad k_j\leq k,  
	\end{equation}
	such that $\sum_{j=1}^{m} k_j= N$. 
	In our case each $S_{[i]}$ is a qubit system. 
	
	One can define the set of genuine uncorrelated states for a given order higher than $k$. In this way, 
	it is possible to define a specific partitioning  considering an integer number $2\leq k \leq N $ and the coarse grained partitioning $\{ \M{H}_{k_1}\otimes\ldots\otimes \M{H}_{k_m} \}$, 
	where each cluster $\M{H}_{k_j}$ includes at most $k$ subsystems \cite{girolami2017}.  
	\begin{defi}[$k$-partite genuine product states]
		It is defined a set of states that has up to $k$ subsystems as  
		\begin{equation}
		P_{k}\coloneqq \left \{\sigma_{N}=\bigotimes_{j=1}^{m}\sigma_{k_{j}}, \sum^{m}_{j=1}k_{j}=N, k=\max\{k_{j}\}\right \},
		\end{equation}
		where $\sigma_{k_{j}}$ is a subsystem of $k_{j}$ particles, this set contains all the sets $P_{k'}$ with $k'<k$, such that  $P_{1}\subset P_{2}...\subset P_{N-1}\subset P_{N}$.
	\end{defi}
	
	In order to calculate the GMC of order higher than $k$, it is used the relative entropy as a pseudo-distance  
	\begin{equation}\label{eq: correlation ktoN}
	S^{k\rightarrow N}(\rho_{N}) = \min_{\sigma \in P_{k}}S(\rho_{N}|| \sigma),
	\end{equation}
	where the minimization is taken over all product states $\sigma = \bigotimes^{m}_{i=1} \sigma_{k_{i}} \in P_{k}$.
	The state $\sigma$ that minimizes $S^{k\rightarrow N}(\rho_{N})$ will be the product of the reduced states of $\rho_N$ \cite{girolami2017, bennett2011, szalay2015}.  Therefore 
	\begin{align}
	\label{eq:dis}
	S^{k\rightarrow N}(\rho_{N})&= S(\rho_{N}||\otimes^{m}_{i=1}\rho_{k_{i}}) \\
	& =\sum_{i=1}^{m}S(\rho_{k_{i}})-S(\rho_{N}). 
	\end{align}
	For states with permutation symmetry  Eq.~\eqref{eq:dis} can be simply written as
	\begin{equation}
	\label{eq:Dis}
	\begin{split}
	S^{k \rightarrow N}(\rho_{N})=&\left \lfloor N/k \right \rfloor S(\rho_{k})+\\
	&(1-\delta_{N\mod k,0})S(\rho_{N\mod k}) - S(\rho_{N}),
	\end{split}
	\end{equation}
	where $\left \lfloor N/k \right \rfloor $ is the floor function, which is the greatest integer less than or equal to $N/k$. The  $\rho_{N\mod k}$ describe the subsystem $S_{N\mod k}$. If we choose $k=1$,
	\begin{equation} \label{eq:totalcorr}
	S^{1 \rightarrow N}(\rho_{N})=N S(\rho_{1}) - S(\rho_{N})
	\end{equation}
	describes the total correlations presented in the system.

	The genuine $k$-partite correlations can be defined as the difference between the correlations of order higher than $k-1\rightarrow N$ and those of order higher than $k\rightarrow N$ 
	\begin{equation}
	\label{eq:Corr}
	S^{k}(\rho_{N})=S^{k-1 \rightarrow N}(\rho_{N})-S^{k \rightarrow N}(\rho_{N}).
	\end{equation}
	Once the correlations of order higher than $k-1$ encapsulates those ones of order higher than $k$, the difference between them returns only the genuine $k$-partite correlations. 
	
	A nice interpretation of the GMC of order $k$ introduced above is that the sum of all GMC gives the total correlation in the system,
	$S^{1 \rightarrow N}(\rho_{N})=\sum_{k=2}^N S^k(\rho_N)$, as can be verified from Eq. (\ref{eq:Corr}).


	\subsection{Lipkin-Meshkov-Glick Model}
	The LMG model, as studied here, is a system composed of $N$ spins $1/2$ fully connected with anisotropy controllable by the parameter $\gamma$ and an external transversal magnetic field $h$ acting on it. This model has first and second order QPTs depending on the values of the control parameters $h$ and $\gamma$. Here, we limit our analysis to the cases in which $\gamma=0.5$ and the external field varies in the range $0\leq h \leq2$. The critical point at $h=1$ signals a second order QPT. Such phase transition has already been studied according to the quantum information theory in Refs. \cite{latorre2005,dusuel2005,dusuel2004,vidal2004,vidal20042,vidal2007}. The Hamiltonian of the LMG is described by \cite{latorre2005,dusuel2004,dusuel2005}, 
	\begin{equation}
	\label{eq:Ham}
	H=-\frac{\lambda}{N} \sum_{i<j} (\sigma_{x}^{i}\sigma_{x}^{j}+\gamma\sigma_{y}^{i}\sigma_{y}^{j})-h\sum_{i=1}^{N}\sigma_{z}^{i},
	\end{equation}
	where $\lambda$ is the ferromagnetic coupling factor, $\lambda=1$ was chosen for the sake of simplicity, and $\sigma_{\alpha}^{k}$ are Pauli matrices with $\alpha=x,y,z$. Employing $J_{\alpha}=\sum_{i=1}^{N}\sigma_{\alpha}^{i}$, the Hamiltonian can be represented by collective spin operators as 
	\begin{equation}
	\label{eq:lip}
	\begin{split}
	H=&-\frac{\lambda}{N}(1+\gamma)(\mathbf{J^2}-J_{z}^{2}-N/2)-2hJ_{z}\\
	&-\frac{\lambda}{2N}(1-\gamma)(J_{-}^{2}+J_{+}^{2}),
	\end{split}
	\end{equation}
	with $\mathbf{J}$ being the total collective angular momentum, $J_{z}$ is its projection along the $z$ direction and $J_{-} $ and $J_{+}$ are collective ladder operators of lowering and raising, respectively. The ground state of the LMG model is a linear combination of Dicke states, which are eigenstates of $\mathbf{J^{2}}$ and $J_{z}$, 
	\begin{equation}
	\begin{split}
	&\mathbf{J^{2}}\ket{J,M}=J(J+1)\ket{J,M},\\
	&J_{z}\ket{J,M}=M\ket{J,M},
	\end{split}
	\end{equation}
	where $J=N/2$ and $M=-N/2, -N/2+1,...., N/2-1, N/2$. Instead of $\ket{J,M}$, it will be used the following representation of Dicke states $\ket{N,n_{e}}$, in which $n_{e}$ is the number of excited spins.  The Dicke states are totally symmetric by permutation of particles and can be represented by
	\begin{equation}
	\label{eq: dicke states}
	\ket{N,n_e}=\frac{1}{\sqrt{\binom{N}{ne}}}\sum_{i}\mathcal{P}_{i}\left(\ket{0}^{N-n_{e}}\otimes\ket{1}^{n_{e}}\right).
	\end{equation}
	The sum is taken over all possible permutations of $n_e$ described by the permutation operator $\mathcal{P}_{i}$ and $\binom{N}{ne}$ is the binomial coefficient required to normalize the Dicke state. Therefore, the ground state of the LMG model is 
	\begin{equation}\label{eq:ground_state}
	\ket{\Psi}=\sum_{n_{e}=0}^{N}P_{n_{e}}\ket{N,n_e},
	\end{equation}
	with $P_{n_{e}}$ being the amplitudes of probability of occurrence of a Dicke state with $n_{e}$ excitations, which ones are obtained from numerical diagonalization of the Hamiltonian in Eq. (\ref{eq:lip}). 
	
	\subsubsection{Quantum Phase Transitions of the LMG Model}
	
	In the Ref. \cite{dusuel2005}, where by mean field approximation was determined the phase diagram of the LMG model, the authors show that for the region of $0\leq h\leq1$ (\textit{broken phase}), the ground state of the system is double degenerated for $\gamma\neq1$, for $\gamma=1.0$ the ground state is infinitely degenerated,  and for $1<h\leq2$ \textit{symmetric phase} the ground state is unique for all $\gamma$. The second order QPT in the LMG model occurs due to the competition between the spins interaction and the effect of the external field $h$ applied over the spin chain. When the external field is strong enough ($h \ge 1$), all the spins begin to align with it so that the state of the system in this phase has no correlations between the spins. For $h=0$ and $\gamma=0$ the ground state of the system is described by a GHZ-like state \cite{latorre2005}. Even though we are using the anisotropy parameter $\gamma=0.5$ for all of our calculations, the ground state is still an approximation of the GHZ-like state, so the spins are correlated. Some previous works in quantum information theory used entanglement \cite{latorre2005,dusuel2005,dusuel2004,vidal2004,vidal20042,vidal2007,barthel2006,caneva2008,cui2008} and others correlations \cite{sarandy2009,ma2008,kwok2008,orus2008,ma2009,leung2012,wang2012} as order parameter to detect the second order phase transition of this model. Furthermore, there are also some previous works in quantum information that make use of FSS \cite{dusuel2004,dusuel2005,ma2008,kwok2008,orus2008,wang2012,cui2008,leung2012} for the calculus of exponents in the LMG model.
	
	
	Since the LMG model is an infinity coordinated system, because all particles interact with all others equally, the system does not have the concepts of length and dimensionality defined \cite{botet1983}, such as made to study finite size scaling in others systems. Then, the number of particles $N$ is the only one variable in the analysis of FSS exponents. Here, we analyse the behavior of the genuine $k$-partite correlations near the second order QPT. 
	The method to extract the exponent that we apply is the following: we take the minimum derivative of $k$-partite correlation and calculate the $k$-partite correlation at this point as function of $N$, thus assuming that relation of correlation and $N$ obeys a power law, we take the logarithm of both variables and we are able to get the exponent relate to the genuine $k$-partite correlation.     
	
	\section{Results}
	\label{sec:res}
	In this section we calculate numerically the genuine $k$-partite correlations across the QPT for some values of $k$. By induction, we conclude that all orders of the genuine multipartite correlations signal the second order QPT at $h=1$. Following the same reasoning, we calculate critical exponents for some genuine $k$-partite correlations and evidence that all exponents are in the range $[-1/2,1/2]$.  
	
	From  Eq.~\eqref{eq:Dis}, we notice that to compute the multipartite correlations of order higher than $k$, $S^{k \rightarrow N}(\rho_{N})$, it is necessary to calculate the reduced density matrix $\rho_k=\text{Tr}_{N/k}\rho_N$ from $\rho_N$. For the ground state, which one is a superposition of Dicke states, as in Eq.~\eqref{eq:ground_state}, we can do the Schmidt decomposition of the Dicke states \cite{stockton2003,toth2007} as  
	\begin{equation}\label{eq:ground_state SD}
	\ket{\Psi}=\sum_{n_{e}=0}^{N}P_{n_{e}}\sum_{l_{e}=0}^{L}\lambda_{l_{e}}\ket{L,l_{e}}\otimes \ket{N-L,n_{e}-l_{e}},
	\end{equation}
	where $\lambda_{l_{e}}=\sqrt{\binom{L}{l_{e}}\binom{N-L}{n_{e}-l_{e}}\binom{N}{n_{e}}^{-1}}$, getting the reduced density matrix from this state and tracing out $\ket{l_{e},L}$ we obtain the reduced density matrix of $k$ spins
	\begin{equation}
	\label{eq:rdm}
	\begin{split}
	\rho_{k}=&\sum_{n'_{e},n_{e}=0}^{N}\sum_{l{e}=0}^{L}\left[\frac{\binom{L}{l_{e}}\sqrt{\binom{N-L}{n_{e}-l_{e}}\binom{N-L}{n'_{e}-l_{e}}}}{\sqrt{\binom{N}{n_{e}}\binom{N}{n'_{e}}}} \right] \\
	&P_{n_{e}}P_{n'_{e}}^{*}\ket{k,n_{e}-l_{e}}\bra{k,n'_{e}-l_{e}},
	\end{split}
	\end{equation}
	where $0\leq n_{e}-l_{e}\leq N-L=k$ and $0\leq n'_{e}-l_{e}\leq N-L=k$. The coefficients $P_{n_{e}}$ depend on the ground state of the LMG Hamiltonian in Eq.~\eqref{eq:lip}.
	
	\begin{figure}[!ht]
		\includegraphics[width=0.48\textwidth]{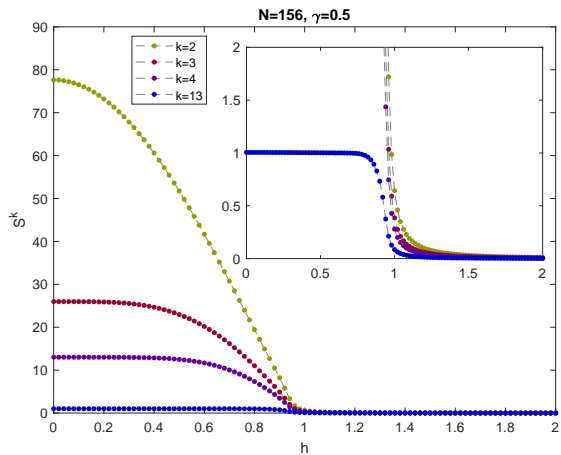}
		\caption{Genuine $k$-partite correlations presented in the ground state of the LMG model for $N=156$, $\gamma=0.5$, and $k=2,3,4,13$ as function of $h$. In the phase in which $h>1$ the spins are aligned to the external field so that they are not correlated. The inset is a zoom of the figure to a better visualization of the behavior of the GMC of order $k=13$ near the phase transition.} \label{fig:1}
	\end{figure}
	 


	\subsection{Genuine $k$-partite correlations in the ground state of the LMG model}

	We recall that the anisotropy parameter is fixed as $\gamma=0.5$ in all numerical analysis of the GMC of order $k$ presented in the ground state of the LMG model. In Fig.~\ref{fig:1} and inset it is possible to observe the behavior of the GMC of orders $k=2,3,4,13$ for $N=156$ spins. For $h=0$ the ground state of the LMG model is approximately a GHZ state \cite{latorre2005}, which implies that the GMC can be described approximately by the expression $S^k=\lceil N/(k-1) \rceil-\lceil N/k \rceil$.  Also, as $k$ increases the genuine $k$-partite correlations decrease and become more resistant to variations of the magnetic field. Such a tendency also appears in Ref \cite{latorre2005} for the calculus of entanglement entropy, where partitions with higher number of particles, up to half of the total number of particles, are more resistant to changes in the field in the region $0\leq h\leq1$. 
	
	\begin{figure}[!ht]
		\includegraphics[width=0.48\textwidth]{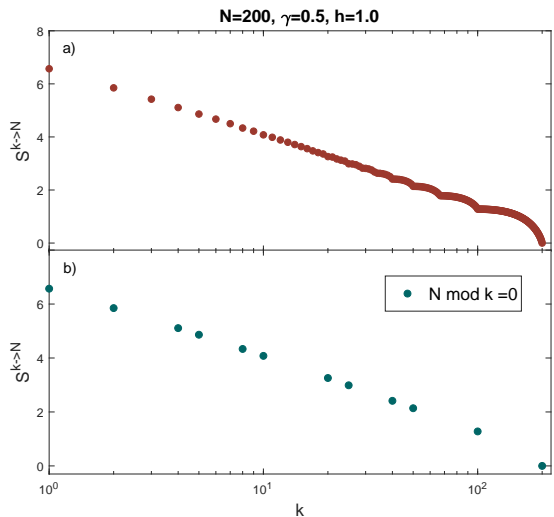}
		\caption{The GMC of order higher than $k$, $S^{k\rightarrow N}$, as function of $k$ for $N=200$, $\gamma=0.5$, and $h=1$. In Fig (a) all integers values of $k$ from $1 \le k \le N$ are considered, while in Fig. (b) only the values of $k$ which satisfy the constraint $N\mod{k}=0$ are take into account.} \label{fig:distance}
	\end{figure}

	Before we start the analysis between the GMC and the second order QPT in the LMG model, we call the attention to the behavior of the distance measure $S^{k\rightarrow N}$ for different values of $k$ and its connection to the total number of spins $N$. The Fig.~\ref{fig:distance} a) shows the distance $S^{k\rightarrow N}$ for $N=200$ and $h=1$ as function of $k$. The distance is monotonically decreasing with the block size $k$. For small values of $k$ the decreasing is smooth, but as $k$ becomes comparable to $N$ abrupt changes occur in form of a  ladder. Such effect was already verified for Dicke states in Ref. \cite{calegari2019} and comes from the floor function in Eq. (\ref{eq:Dis}), or equivalently, there is an unpaired block of $k'$ particles with $k'<k$. 
	On the other hand, in Fig.~\ref{fig:distance} b) we also show the dependence of $S^{k\rightarrow N}$ with $k$, but with the constraint $N \mod{k}=0$. As can be seen, this is enough to remove the role played by the floor function and consequently removing the ladder behavior. This kind of imposition allow us to get the critical exponents straightforwardly, since the unpaired blocks cause abrupt changes in the GMC of order $k$.
	\begin{figure}[!h]
		\includegraphics[width=0.48\textwidth]{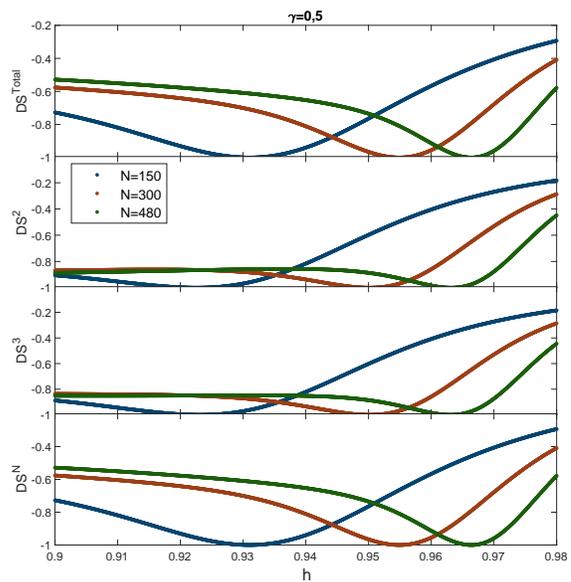}
		\caption{The first order derivative of the total correlations ($k=1$) and GMC of orders $k=2,3,N$ as function of the external control parameter $h$. The minimum value of the derivatives occurs for values of $h$ closer to $h_c=1$ as $N$ increase.} \label{fig:derivative}
	\end{figure}
	
	\subsection{Quantum phase transition and genuine $k$-partite correlations}
	
	Our first task is to show that all GMC of order $k$ ($1 \le k \le N$) are able to signal the second order QPT. For the sake of simplicity we included the value $k=1$, which means the total correlations, see Eq. (\ref{eq:totalcorr}). In order to verify that the GMC signal the QPT we calculate the first derivative of $S^k$ with respect to the parameter $h$ and show that in the thermodynamic limit it is non analytic for some value of $h$, named critical parameter $h_c$. This behavior is shown in Fig. \ref{fig:derivative}, where we conclude that the minimum value of $d S^{k}/dh$ (for $k=1,2,3,N$) occurs for some value of the control parameter $h_{min}$ which tends to $h_c=1$ for $N \rightarrow \infty$.
	 
\begin{figure}[!ht]
		\includegraphics[width=0.48\textwidth]{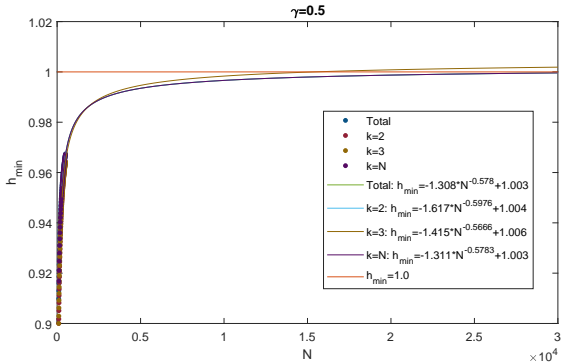}
		\caption{Values of the control parameter $h$ in which the first order derivative of the total correlations and GMC of orders $k=2,3,N$ are minimum as function of the number of spins in the system $N$. The solid lines are logarithm fit of the points to show that in the thermodynamic limit ($N\rightarrow \infty$ $h_{min}) \rightarrow h_c=1$. The constraint $N\mod{6}=0$ has been imposed for the curves with $k=2,3$. \label{fig:first_derivative_therm}}
	\end{figure}

	 This result is in agreement with Fig.~\ref{fig:1} inset, where the GMC of higher orders disappear faster after the phase transition $h_c=1$ and with Fig. \ref{fig:first_derivative_therm} which explores the thermodynamic limit. We observe that the same procedure has been performed for other values of $k$ not reported here, but which ones corroborate the conclusion presented above. Therefore, the GMC of order $k$ are able to signal the already known second order QPT in the LMG model \cite{dusuel2005}. 
	
	\subsection{Finite size scaling exponent analysis of genuine $k$-partite correlations}
	
\begin{figure}[!h]
	\includegraphics[width=0.48\textwidth]{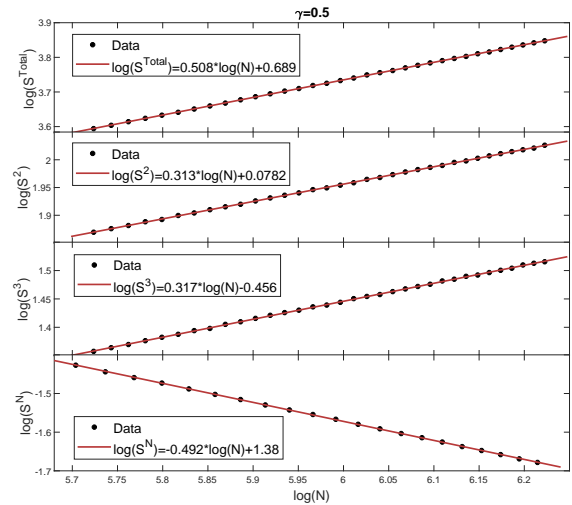}
	\caption{The critical exponents are obtained from the angular coefficients of the line for the total correlations and for the CGM of order $k=2,3,N$. For $k=2,3$ the constraint $N\mod{6}=0$ has been imposed.}  \label{fig:critical_expoent}
	\end{figure}

	The application of the finite size scaling method to find the exponents that govern the behavior of the order parameter near the transition point can be made using different strategies. Here, to compute the exponent of the genuine $k$-partite correlations we take the minimum point $h_{min}$ of the first derivative of the correlation as function of the number of particles $N$. Then, the critical exponent $\alpha$ is obtained from the function $S^k(h_{min})=AN^{\alpha}$, where $A$ is some constant.  As the computational cost for the numerical calculations increases rapidly with the number of particles $N$, we calculate the critical exponents extending $N$ until $500$ spins, i.e., our thermodynamic limit in practice. The values of the exponents can vary a little, becoming smaller with the increasing of $N$ and when ruling out the smaller values of $N$ in the graph. The strategy used here to obtain the critical exponents considers only the values of $N$ in which $N \mod k=0$ and $N \mod (k-1)=0$. As mentioned before, the advantage of this procedure is that it avoids sudden changes in the values of the GMC of order $k$ and consequently enable to obtain a well defined critical exponent. However, as $k$ increases fewer points remain in the graphs to calculate the exponents. This method is applied to obtain the critical exponents of the total correlations and for correlations of order $k=2,3,N$, as shown in Fig.~\ref{fig:critical_expoent}. Additional results for $k=4,5,N/4,N/2$ are summarized in Table \ref{Tab:first}.
		\begin{table}[h!]
		\begin{center}
			\begin{tabular}{ | l | c |  }
				\hline
				$k$ & Critical exponent \\ \hline
				$1$ & $0.508 \pm 0.001$ \\ \hline
				$2$ & $0.313 \pm 0.001$ \\ \hline
				$3$ & $0.317 \pm 0.002$ \\ \hline
				$4$ & $0.333 \pm 0.004$ \\ \hline
				$5$ & $0.350 \pm 0.003$ \\ \hline
				$N/4$ & $-0.377 \pm 0.002$ \\ \hline
				$N/2$ & $-0.4540 \pm 0.0007$ \\ \hline
				$N$ & $-0.492 \pm 0.003$ \\ \hline
			\end{tabular}
			\caption{Critical exponents for different values of $k$ of the GMC $S^k$ vs. $h$ across the second order QPT in the LMG model.} \label{Tab:first}
		\end{center}
	\end{table}
	Although we did not test all possible values of the critical exponents, it seems that they are confined in the interval $[-1/2,1/2]$. In literature we have found the critical exponents for bipartite correlations only, being $1/3$ for entanglement entropy \cite{latorre2005} and concurrence \cite{vidal20042}. From Table \ref{Tab:first} we notice that for partitions of fixed size ($k=1,2,3,4,5$), the critical exponent is positive, but when the partition size increases with the number of spins ($k=N/4,N/2,N$), the critical exponent becomes negative. Also, the GMC across the quantum phase transition diminish faster for higher orders of $k$ when it depends on $N$. In the particular case of the GMC of order $N$ it goes to zero at a rate greater than for the other exponents. If we analyse the GMC of order $k$ per particle in the thermodynamic limit across the second order QPT, $\lim_{N\rightarrow \infty} S^k(\rho_N)/N =0$, it becomes null, as expected for the classical world.

	\section{Conclusions and Perspectives}\label{sec:conclusion}
	
	We analysed the genuine multipartite correlations in the Lipkin-Meshkov-Glick model according to the measure introduced by Girolami and coworkers \cite{girolami2017}.  Within this framework we were able to calculate the genuine $k$-partite correlations and show they behavior for different partition sizes. Also, we verify that the genuine $k$-partite correlations signal the second order quantum phase transition in the LMG model. Furthermore, we obtained the critical exponents through finite-size scaling analysis for the total correlations and genuine multipartite correlations of order $k=2,3,4,5,N/4,N/2,N$. From that we observed that the genuine multipartite correlations go to zero across the second order quantum phase transition in the thermodynamic limit when the partition size $k$ increases with the number of particles of the system.   
	
	As perspective for future works, it would be interesting to certify if the critical exponents of the genuine multipartite correlations of order $k$ are confined in the range $[-1/2,1/2]$ for all $k$. Likewise, the analysis of weaving, a measure of correlation also proposed by Girolami et. al \cite{girolami2017} which is the sum of all genuine multipartite correlations, to see if it is possible to gain additional information on the scalability of the genuine multipartite correlations across the quantum phase transition.     
	
	\begin{acknowledgments}
		The authors acknowledge financial support from
		the Brazilian funding agencies Coordenação
		de Aperfeiçoamento de Pessoal de Nível Superior
		(CAPES), Conselho Nacional de Desenvolvimento
		Científico e Tecnológico (CNPq), Fundação de Amparo à Pesquisa e Inovação do Estado de Santa Catarina (FAPESC) and Instituto
		Nacional de Ciência e Tecnologia de Informação
		Quântica (CNPq INCT-IQ (465469/2014-0)).
	\end{acknowledgments}
	
	\bibliography{references}
	
\end{document}